\documentclass[letterpaper, 11 pt, conference, onecolumn]{article}

\usepackage{graphics} 
\usepackage{epsfig} 
\usepackage{mathptmx} 
\usepackage{times} 
\usepackage{amsmath} 
\usepackage{amssymb}  
\usepackage{hyperref}
\usepackage[font=small]{caption}
\usepackage{pdfpages}

\title{\LARGE \bf
CID Models on Real-world Social Networks and Goodness of Fit Measurements
}
\author{Jun Hee Kim \\ \texttt{\small junheek1@andrew.cmu.edu} \and Eun Kyung Kwon \\ \texttt{\small eunkyunk@andrew.cmu.edu} \and Qian Sha \\ \texttt{\small qsha@andrew.cmu.edu} \and Brian Junker \\ \texttt{\small brian@stat.cmu.edu} \and Tracy Sweet \\ \texttt{\small tsweet@umd.edu}}
\date{}

\begin{document}

\maketitle
\thispagestyle{plain}
\pagestyle{plain}
\footskip = 30pt
\hoffset = 0pt
\paperwidth = 597pt

\begin{abstract}
\normalfont Assessing the model fit quality of statistical models for network data is an ongoing and under-examined topic in statistical network analysis. Traditional metrics for evaluating model fit on tabular data such as the Bayesian Information Criterion are not suitable for models specialized for network data. We propose a novel self-developed goodness of fit (GOF) measure, the ``stratified-sampling cross-validation'' (SCV) metric, that uses a procedure similar to traditional cross-validation via stratified-sampling to select dyads in the network's adjacency matrix to be removed. SCV is capable of intuitively expressing different models' ability to predict on missing dyads. Using SCV on real-world social networks, we identify the appropriate statistical models for different network structures and generalize such patterns. In particular, we focus on conditionally independent dyad (CID) models such as the Erdos Renyi model, the stochastic block model, the sender-receiver model, and the latent space model.
\end{abstract}

\section{Introduction}
\noindent
Social networks are ubiquitous in our society, and they represent important interactions among individuals. The interdependency of entities makes social networks compelling, yet complicated and difficult to analyze. Many statistical models for network data have been proposed and their mathematical properties have been examined. For example, the Erdos-Renyi model assumes that every pair of nodes has an equal probability of forming an edge, while the sender-receiver model assumes each node has its own probability of sending an edge and that of receiving an edge. However, whether these models actually fit well on real-world network datasets remains an unsolved issue. In fact, quantitative goodness of fit (GOF) measures for statistical models on network data have not been considered to a great extent. Current GOF metrics on models for non-network data fail to capture features of network data. For example, the Akaike Information Criterion and the Bayesian Information Criterion penalize model complexity, but not only is it difficult to settle on suitable criteria for complexity but also undesired to penalize for an essential feature of social network data. Therefore, in this paper, we identify appropriate conditionally independent dyad (CID) models for different network structures and propose a metric of GOF assessment to evaluate the developed statistical models for network data.

This paper first introduces, in Section 2, basic concepts and properties of graphs as a data structure, specifically those frequently used throughout this study. In Section 3, four CID models for network data analysis are discussed: the Erdos-Renyi model, the stochastic block model, the sender-receiver model, and the latent space model. Section 4 describes our initial attempt of developing a GOF metric using cross-validation and its limitations, followed by our remedy, the ``stratified-sampling cross-validation'' (SCV) metric. Section 5 introduces a set of real-world social networks that were analyzed throughout the study. Section 6 discusses the SCV accuracy results for each network, and Section 7 reports the fit of the best model chosen according to the results in Section 6. Section 8 generalizes the conclusions of this study. Lastly, we propose potential future work in Section 9 considering the limitations in our research.

\section{Graph Terminology}
\noindent
In this section, we introduce a set of basic graph terminologies to help understand the applications mentioned in the following sections. A network is mathematically represented as a graph $G = (V, E)$, where $V$ is the set of vertices (nodes or actors) and $E$ is the set of edges (links or ties) that are either directed or undirected. In social networks, entities are usually represented as nodes, and a social relationship between any two entities is denoted as an edge connecting those two corresponding nodes. Note that an undirected edge $i-j$ can be viewed as two directed edges $i \rightarrow j$ and $j \rightarrow i$. Furthermore, the adjacency matrix, also known as incidence matrix or sociomatrix, is often used to represent a finite graph. It is defined as an $|V| x |V|$ matrix $A$ such that $A_{ij} = 1$ if there exists an edge from node $i$ to node $j$ and $0$ otherwise, where $A_{ij}$ denotes the $i^{th}$ row, $j^{th}$ column entry of matrix $A$. A diagonal entry represents whether a node has an edge to itself, which we assume to be not true in the scope of our study.

Reciprocity and density are two commonly used descriptive statistics for networks. Reciprocity is a measure of the likelihood that nodes in a directed network are mutually linked. The reciprocity of a network is computed as:

\begin{equation}
\text{Reciprocity} = \frac{\sum_{i=1}^{n} \sum_{j=1}^{n} A_{ij} A_{ji}}{\sum_{i=1}^{n} \sum_{j=1}^{n} A_{ij}}
\end{equation}
where $n = |V|$. Note that in this paper, if no additional description is provided, $n$ always denotes $|V|$, the number of nodes in the network. Reciprocity is frequently used to study complex networks as it gives a sense of how likely the relationships are mutual rather than one-directional. By definition, an undirected graph has reciprocity of 1 since a pair of nodes having an undirected edge is equivalent to having two directed edges in both directions and thus the relationship is mutual.

Density measures how close the number of edges is to the maximum possible number of edges, which is determined by the number of nodes. The formula for graph density is:

\begin{equation}
\text{Density} = \left\{
        \begin{array}{ll}
            \frac{2 |E|}{|V|(|V|-1)} & \quad \text{undirected graph} \\
            \frac{|E|}{|V|(|V|-1)} & \quad \text{directed graph}
        \end{array}
    \right.
\end{equation}
The higher the density of a network, the more the nodes are likely to be linked to one another. With this measure we can not only assess how closely knitted a network is, but also easily compare networks against each other or at different regions within a single network.

\section{Conditionally Independent Dyad (CID) Models}
\noindent
In this section, we examine a family of statistical models specifically developed for network data, namely the conditionally independent dyad (CID) models where dyads, which are random variables for each pair of distinct nodes on whether or not there is an edge, are assumed to be conditionally independent given the model parameters. We denote the probability that an edge from node $i$ to node $j$ exists by $p_{ij}$. Again, throughout this study we assume that a node cannot have an edge to itself. That is: for any node $i$, $p_{ii} = 0$. Then the likelihood is: 

\begin{equation}
P(A=A^{(obs)}| \theta) = \left\{
        \begin{array}{ll}
            \prod_{i < j} p_{ij}^{A_{ij}^{(obs)}} (1-p_{ij})^{1-A_{ij}^{(obs)}}
 & \quad \text{undirected graph} \\
            \prod_{i \ne j} p_{ij}^{A_{ij}^{(obs)}} (1-p_{ij})^{1-A_{ij}^{(obs)}}
 & \quad \text{directed graph}
        \end{array}
    \right.
\end{equation}
where $A^{(obs)}$ is the observed adjacency matrix, and $\theta$ denotes the set of model parameters (so each $p_{ij}$ depends on $\theta$).

\subsection{Erdos-Renyi Model}
\noindent
The Erdos-Renyi model (ER)$^{1}$ is the simplest CID model. The model assumes that each pair of distinct nodes in a network has equal probability $p$ of having an edge. That is: for every pair of nodes $i$ and $j$ where $i \ne j$, 
\begin{equation}
p_{ij} = p
\end{equation}
When we fit the Erdos-Renyi model to network data, the optimal value of $p$, the only parameter, is estimated.

\subsection{Stochastic Block Model}
\noindent
The stochastic block model (SBM)$^{2}$ is commonly used for detecting block structures in networks. The SBM is defined by three components:

\begin{enumerate}
\item A number $k$, representing the number of blocks to be fitted
\item An $n$-dimensional vector $\vec{z}$, where the $i^{th}$ component $z_i$ denotes block assignment of node $i$
\item A $k \times k$ stochastic block matrix $M$, where $M_{ij}$ denotes the probability that there is an edge from a node assigned to block $i$ to a node assigned to block $j$
\end{enumerate}
where $k$ is a hyperparameter that must be pre-specified, while $\vec{z}$ and $M$ are model parameters to be estimated.

The diagonal entries of $M$ are probabilities that nodes assigned to the same block have an edge, while the non-diagonal entries are probabilities that nodes assigned to different blocks have an edge. In most applications of the SBM, especially for community detection, it is desired that entries on the diagonal of $M$ are large and those on the off-diagonal blocks are small. In this case, nodes within the same block are more likely to be connected compared to those in different blocks. However, there are exceptions (opposite case) where nodes in the same block are not expected to be connected while those in different blocks are. In the scenario of describing marriage relationship between a group of men and a group of women, if each group is represented as a block in an SBM (so total $2$ blocks), then the opposite case holds. What the \textsf{CIDnetworks} R package does when fitting an SBM is to first, for each node, compute the probability that this node should be assigned to each of the $k$ blocks (so total $k$ probabilities that sum to $1$), and then select the block with the highest probability as the final block assignment for that node.

\subsection{Sender-Receiver Model}
\noindent
The sender-receiver model (SR), which is a special case of the p1 models introduced in Holland \& Leinhardt (1981)$^{3}$, takes into consideration each node's tendency to send edges towards others and also that to receive edges from others. In social network context, these can be though of as how outgoing an entity is and how popular to other people an entity is, respectively. The model has two parameters associated to each node $i$: $\beta_{i}^{(send)}$ and $\beta_{i}^{(receive)}$, which represent how likely this node sends edges to others and receive from others, respectively. The model also has an intercept parameter $\beta_{0}$, so there are total $2n + 1$ model parameters to be estimated. There are two versions of the edge probability formula for the SR model:

\begin{equation}
p_{ij} = 
\left\{
        \begin{array}{ll}
            \Phi(\beta_{0} + \beta_{i}^{(send)}+\beta_{j}^{(receive)}) & \text{probit version} \\
            \frac{\text{exp}\{\beta_{0} + \beta_{i}^{(send)} + \beta_{j}^{(receive)}\}}{1 + \text{exp}\{\beta_{0} + \beta_{i}^{(send)} + \beta_{j}^{(receive)}\}} & \text{logistic version}
        \end{array}
    \right.
\end{equation}
where $\Phi(\cdot)$ denotes the standard Gaussian CDF.

\subsection{Latent Space Model}
\noindent
In some networks, the probability of an edge between two nodes may increase proportional to some measure of similarity. Thus, a subset of entities with high number of edges between them may be indicative of a group of nodes who have adjacent positions in the ``social space'' of characteristics. 

The latent space model (LSM)$^{4}$ takes account of this factor, assigning each node $i$ to a position in the Euclidean space $z(i)$. The edge probability between nodes $i$ and $j$ depends on their positions in the (unobserved) latent space, and the potential strength of a dyad decreases proportionally to the distance between $z(i)$ and $z(j)$. The LSM also has two versions of edge probability formulas:

\begin{equation}
p_{ij} = \begin{cases} \Phi(\mu+X_{ij}+U_{ij}) &\mbox{probit version} \\
\frac{\textrm{exp}(\mu+X_{ij}+U_{ij})}{1+\textrm{exp}(\mu+X_{ij}+U_{ij})} &\mbox{logistic version}
\end{cases}
\end{equation}
where $X_{ij}$ denotes covariates, $U_{ij} = -||z(i)-z(j)||$, and again $\Phi(\cdot)$ denotes the standard Gaussian CDF. 

An advantage of this method is that it can provide visual (if the dimension of the latent space is $\le 3$) and interpretable model-based spatial representations. Note that the latent space's dimension is a hyperparameter which should be pre-specified.

\subsection{Suitability of Cross-validation for CID Models}
\noindent
From CID models, we have an advantage of learning about how nodes relate to one another. For example, entity relations are modelled probabilistically in an LSM. That is, the observation of an edge from node $i$ to node $j$ and also from node $j$ to node $k$ suggests that nodes $i$ and $k$ are not too far apart in the social space, making them more likely to have an edge. 

Likewise, the interdependence between edges is implicit in CID models. If a subset of edges in the graph are missing at random, we do not need further imputation to estimate the model parameters. Again referring to the LSM, we assume that each node $i$ has position $z(i)$ on the unobserved latent space. The edges in the network are presumed to be conditionally independent given the model parameters (latent space position of each node), and the probability of an edge connecting two entities can be modelled as a function of their positions. Without any additional computation, we can utilize this function to predict the missing entries in the adjacency matrix of a graph.

For this reason, cross-validation or out-of-sample prediction can be considered a natural approach to assess GOF for CID models. There are some previous work that introduce cross-validation methods for statistical models for network data. For example, Dabbs (2016)$^{5}$ introduces latinCV that assigns each edge indicator, or dyad, to one of the $K$ folds, and then run the traditional $K$-fold cross-validation, where for each fold, that fold acts as the test set and the others are used for training. The fold assignment in latinCV is done by:
\begin{enumerate}
\item Construct a fixed $n \times n$ fold assignment matrix where each row and column has an equal number of occurrences of each fold.
\item Then, permute the rows and columns randomly to get the finalized fold assignment matrix.
\end{enumerate}

Under this motivation, we chose to base our GOF metric on a cross-validation approach, introduced in the following section.

\section{GOF Measure: Stratified-sampling Cross-validation (SCV)}
\noindent
In this section, we propose a self-developed goodness of fit measure for statistical models on network data. As previously mentioned, defining a criterion for model fit on network data is a difficult problem. 

First, we discuss a self-developed procedure that is similar to traditional $K$-fold cross-validation and its shortcomings. In this method, around $20\%$ of the entries of the adjacency matrix are randomly sampled and set to be missing entries (NA). Then, the model is fit on the remaining entries and estimates of the model parameters are obtained. Based on the probabilistic output of the fitted model, for each node, prediction on whether or not there should be an edge is performed using the edge probability. More specifically, for each missing entry, we compute the estimated edge probability of the corresponding pair of nodes using the parameter estimates and formula according to the model, and predict by running a Bernoulli trial using that probability. This 3-step procedure of deleting, fitting, and predicting is repeated $K$ times, and then the model's prediction accuracy is computed as the average of the $K$ prediction accuracies.

This approach, which is similar to latinCV but does not necessarily assign each dyad to exactly one fold, is simple and intuitive, but it had a limitation. Since the majority of the collected data and real-world social networks in general are sparse, most entries of their adjacency matrices were $0$'s. To take account of this fact we compared the previously-mentioned prediction to zero imputation which simply always predicts as $0$, and observed that the latter yields a much higher prediction accuracy. One problem is that an extremely high proportion of the deleted entries were $0$'s, so a model fitted on an extremely sparse (sub)network inevitably does well in the prediction task. Another issue was that the density of the subnetwork is not reasonably similar to that of the original network since the sampled entries are randomly selected from a bucket where $0$'s are dominant over $1$'s. Hence the model fitted on the subnetwork is not a good approximation, or simulation, of the model being fitted on the original network.

Such limitations motivate our ``stratified-sampling cross-validation'' (SCV) metric, which incorporates an alteration in the sampling procedure. The main objective is to retain a balanced proportion of the $0$'s and $1$'s in the sample, so that density is reasonably preserved even for the subnetwork. For every iteration, we randomly select an equal proportion of edges (i.e. $1$'s) and non-edges (i.e. $0'$s) from the adjacency matrix, using approximately $20\%$ in our simulation. Again, the sampled entries are set to be missing values, and the model is fit on the remaining dyads. Prediction is performed in the same way as before of running a Bernoulli trial using the estimated edge probability for the corresponding pair of nodes. Using this method, the dominance in overall prediction accuracy of zero imputation was moderately controlled for. The procedure of SCV is shown in Figure 1.

To examine the large contrast between prediction on edges compared to non-edges, we recorded results separately for the two cases. Also, due to the randomness involved in the process of sampling the entries and fitting the models, the ranking of accuracies across models can differ among iterations. Therefore, we designed our metric in such a way that each trial runs $10$ iterations of this process, and the average of the $10$ accuracy values is computed as the final result of that trial. We use this prediction accuracy as a representative measure of how good the model fits the network. The results shown hereafter are based on five such trials, which contain total $50$ iterations.

\begin{figure}[h!]
\centering
  \includegraphics[width=8cm, height=12cm]{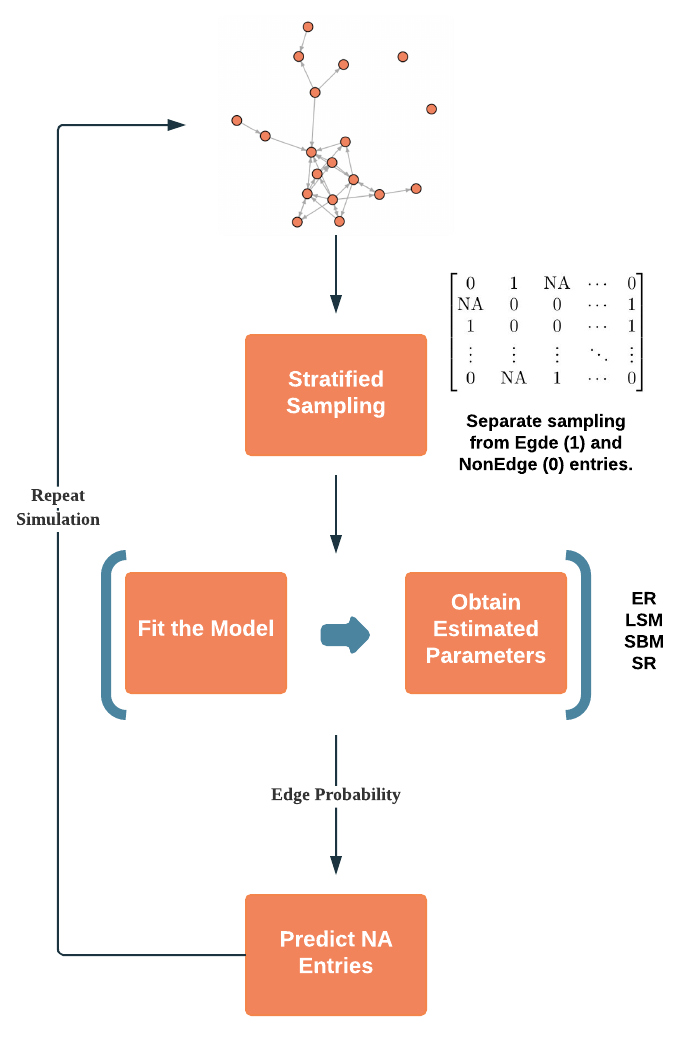}
  \caption{Stratified-sampling Cross-validation (SCV) Procedure}
\end{figure}

\section{Network Data Description}
\noindent
In the following section, we introduce the six real-world social networks analyzed in this study. Table 1 contains the summary statistics for each of the six networks. Three out of the six networks have reciprocity of $1$ and thus are undirected, while the other three are directed. Most of the networks are extremely sparse, and the Freshmen Network is the only graph that has density not below $0.15$. This feature is not surprising since the majority of real-world social networks have very low density. The following subsections report the contexts and exploratory data analyses of the networks.

\begin{table}[htbp]
\centering
\begin{tabular}{ccccc}
\hline
Network & Number of Nodes & Number of Edges & Density & Reciprocity \\
\hline
\hline
Highschool Network & 70 & 366 & 0.076 & 0.503\\
Karate Club Network & 34 & 78 & 0.139 & 1 \\
Oxford Preschoolers Network & 19 & 41 & 0.120 & 0.195 \\
Freshmen Network & 29 &282 & 0.347 & 0.567 \\
Dolphins Network & 62 & 159 & 0.084 & 1 \\ 
Twitter Retweet Network & 96 & 117 & 0.026 & 1 \\
\hline
\end{tabular}
\caption{Summary Statistics of Each Social Network}
\end{table}

\subsection{Highschool Network}
\noindent
The Highschool Network$^{6}$ describes friendships among boys in a small high school at Illinois. Each of the 70 boys was asked in Fall 1957 and then later in Spring 1958 to report the boys whom he considers his friends. For each of the directed edges, the weight is $1$ if that report was made only once, and $2$ if that report was made in both surveys. This dataset does not specify, for the edges with weight $1$, whether the only-one time decision was in the first or the second report.

Figure 2 shows a plot of the network. Note that among the 366 edges, 226 edges have weight $1$, and the other 140 have weight $2$, so it is relatively rare for a relationship to exist in both surveys: that is, to last longer. Moreover, the network has a triangular structure in such a way that visually each node can be seen as being in one of the three vertices of the triangle.

\begin{figure}[h!]
\centering
  \includegraphics[width=6.5cm, height=6.5cm]{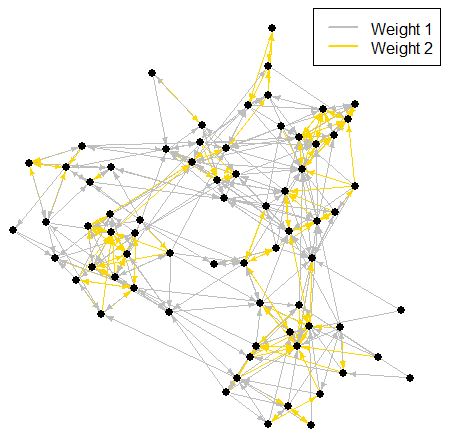}
  \caption{Highschool Network}
\end{figure}
In the analyses discussed in the following sections, the edges of this network are treated as equally-weighted. That is, this network was assumed to be non-weighted.

\subsection{Karate Club Network}
\noindent
The Karate Club Network data$^{7}$ were collected by Wayne Zachary in 1977. They describe an absence or a presence of friendship outside of the university karate club activities between each pair of members. The 34 nodes represent the members of this karate club, and the 78 non-weighted edges represent the friendship between the corresponding two members.

By the nature of this network which aims to determine the existence of mutual friendship, each edge in this network is undirected. Figure 3 shows a plot of this network.

\begin{figure}[h!]
\centering
  \includegraphics[width=6cm, height=6cm]{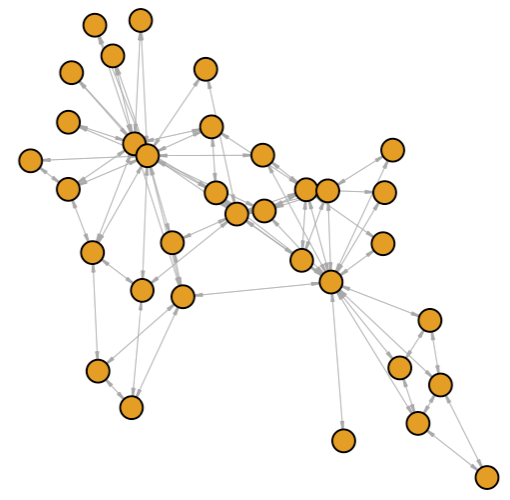}
  \caption{Karate Club Network}
\end{figure}

\subsection{Oxford Preschoolers Network}
\noindent
The Oxford Preschoolers Network$^{8}$ was obtained from a series of ethological observations over preschool children in Oxford, England. These data were derived from observing a group of 19 boys competing for a toy. Each competition occurred between two individuals, and results recorded a ``loser'' and a ``winner'' in each incidence, where the ``winner'' won the toy. This network contains multiple occurrences of such competitions. Every node is labelled by the corresponding child's initials. Each of the 41 directed edges represents a dominance relationship between the two children who competed for a toy and has a weight showing how many times the same incidence (the competition between the same two kids with the same winner) happened. In the following sections, we assume the edges are not weighted.

The reciprocity of the network is also very low (see Table 1), meaning that relationships described in this network are barely mutual, so it is hardly likely that a child $i$ who has lost to another child $j$ in a competition has beat $j$ in a different competition. Hence, we believe the hierarchical structure for dominance is highly consolidated in this society; ``winners'' tended to remain their status and this relationship was not likely to be overturned. In Figure 4, we can also observe that some boys, NP and RC, decided not to engage in the competition at all.

\begin{figure}[h!]
\centering
  \includegraphics[width=6cm, height=6cm]{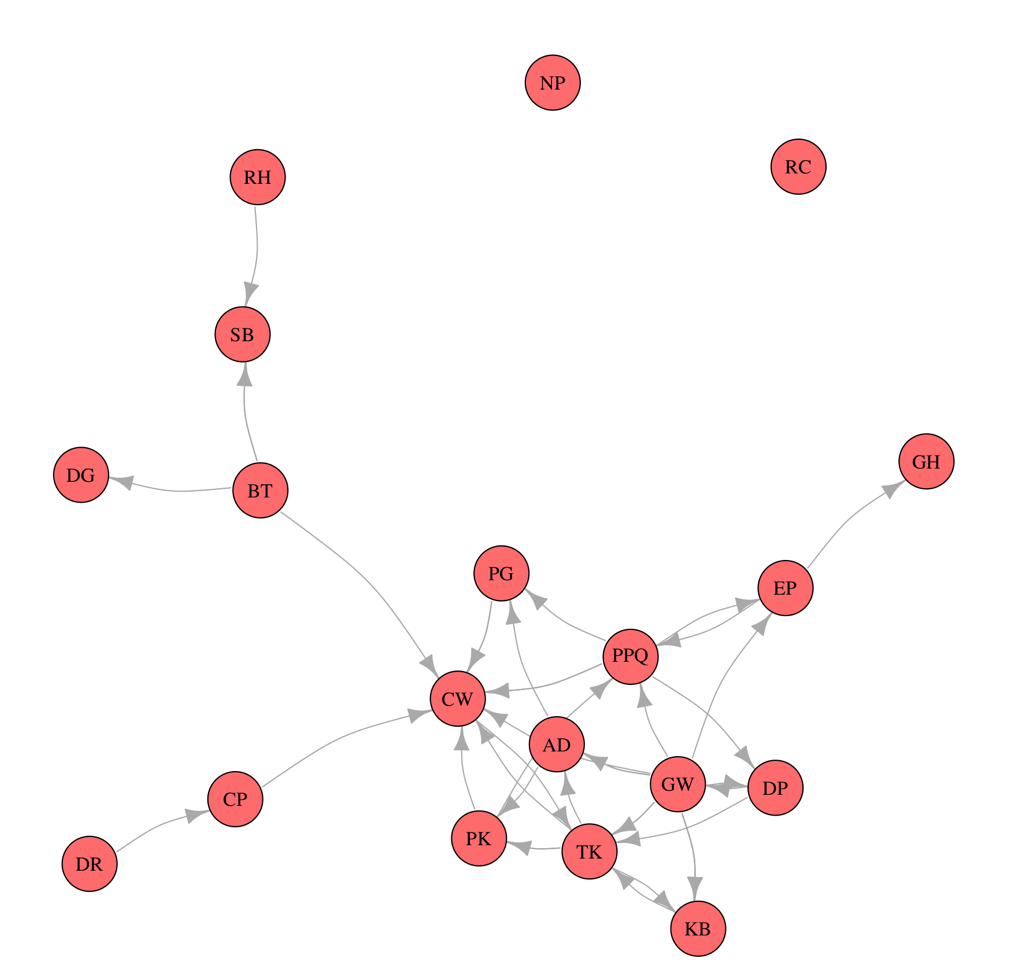}
  \caption{Oxford Preschoolers Network}
\end{figure}

\subsection{Freshmen Network}
\noindent
The Freshmen Network$^{9}$ describes friendship relationships among freshmen studying Sociology in the University of Groningen during 1998-1999. These social network data contain information about 34 freshmen and how intimate each of them considers the other 33 colleagues. 

Each directed edge has weight defined as (When collecting the data, each student reported each relationship by choosing one of these five): $1$ for ``Best friendship'', $2$ for ``Friendship'', $3$ for ``Friendly relationship'', $4$ for ``Neutral relationship'', and $5$ for ``Troubled relationship''.

Note that edges with weight $4$ or $5$ indicate non-intimate relationships that are neutral or hostile. So we only analyze the network only containing edges with weight $1$, $2$, or $3$, which all indicate intimacy of different levels. In this paper, the term ``Freshmen Network'' (including those in Table 1 and Figure 5) refers to this subnetwork only containing edges with one of those three weight values, and not the original network. Figure 5 shows a plot of the network. In the following analyses on this network, the edges are treated as equally-weighted. 

\begin{figure}[h!]
\centering
  \includegraphics[width=8.5cm, height=5.5cm]{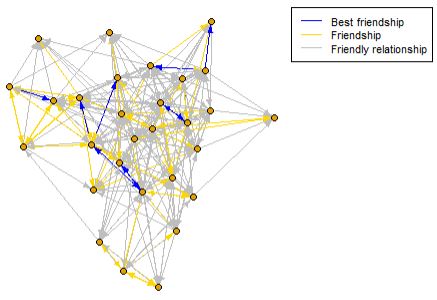}
  \caption{Freshmen Network}
\end{figure}

\subsection{Dolphins Network}
\noindent
The Dolphins Network$^{10}$ describes the associations among 62 bottlenose dolphins. Each of the undirected and unweighted edges indicates frequent associations between the corresponding two dolphins, where for a pair of dolphins $a$ and $b$, the strength of their association is depicted by the half-weight index (HWI): 
\begin{equation}
\text{HWI}(a,b) = \frac{X}{X+0.5\times(Y_{a}+Y_{b})}
\end{equation}
where $X$ is the number of schools where dolphins $a$ and $b$ were seen together, $Y_a$ is the number of schools where dolphin $a$ was sighted but not dolphin $b$, and $Y_b$ is the number of schools where dolphin $b$ was sighted but not dolphin $a$. More specifically, whether or not a pair of dolphins has an edge was determined by the following $3$-step process$^{11}$:

\begin{enumerate}
\item Randomly permute individuals within schools ($20000$ times), keeping the school size and the number of times each individual was seen constant.
\item After each permutation the HWI for each pair is calculated, the observed HWI is compared to the $20000$ expected HWI values.
\item If more than $95\%$ of the expected HWI values are smaller than the observed HWI, the pair of dolphins was defined as a preferred companionship and has an undirected edge. Otherwise, the pair does not have an edge.
\end{enumerate}

Figure 6 shows a visualization of this network, in which the node size reflects the degree of each node in this network graph.

\begin{figure}[h!]
\centering
  \includegraphics[width=6cm, height=6cm]{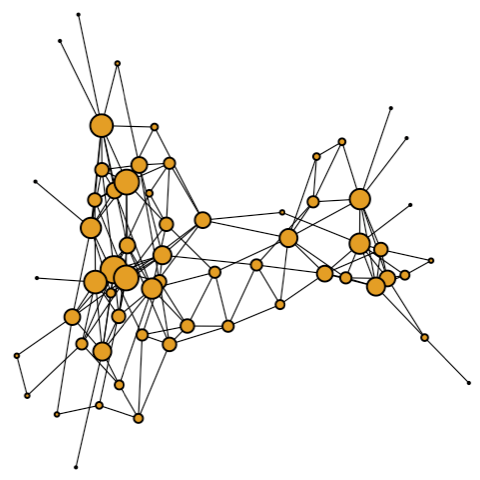}
  \caption{Dolphins Network}
\end{figure}

\subsection{Twitter Retweet Network}
\noindent
The Twitter Retweet Network$^{12}$ contains information about 96 Twitter users and 117 retweets. The network data were collected over social and political hashtags on a social networking site, Twitter. Each edge in this graph is undirected and unweighted, indicating that the corresponding two users have retweeted each other on posts. 

One significant structural feature in the network are cliques (Figure 7), which are groups of users that have all mutually retweeted each other. They potentially represent an interest cartel, which can be important information for applications like personalized recommendations.

\begin{figure}[h!]
\centering
  \includegraphics[width=6cm, height=5.5cm]{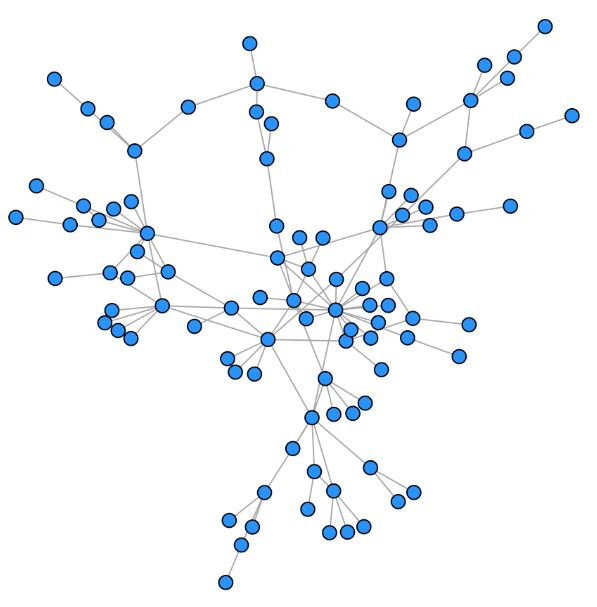}
  \caption{Twitter Retweet Network}
\end{figure}

\section{Stratified-sampling Cross-validation Results}
\noindent
For each of the networks discussed in Section 5, four CID models were tested in SCV: the Erdos-Renyi model (ER), the stochastic block model (SBM), the sender-receiver model (SR) and the latent space model (LSM). Each fit had five SCV trials, where each trial runs $10$ iterations of the procedure of sampling, fitting, and predicting. As mentioned before, the ER model has only one parameter $p$ and hence is considered the simplest CID model. Therefore, it is expected that non-ER models, if appropriate for a network, have better prediction accuracy than the ER model's since they should identify certain structures or patterns in the network that the ER model is not capable of capturing.

Figure 8 shows the SCV results for all six networks. Note that the red boxplots represent prediction accuracies on non-edges, and the blue boxplots describe those on edges. We observe a big difference in the prediction accuracies on edges from those on non-edges. Overall, there was a tendency of this discrepancy to decrease as the network is more dense. Considering the high sparsity of the examined networks, we find it more meaningful to examine prediction on edges. Moreover, our primary interest lies on the existence of interaction between nodes to understand how entities are related to each other. Therefore, we focus on the prediction accuracy on edges (i.e. $1$'s in the adjacency matrix) when choosing the final model to fit for each network.

\begin{figure}[h!]
\centering
  \includegraphics[width=16cm, height=7.5cm]{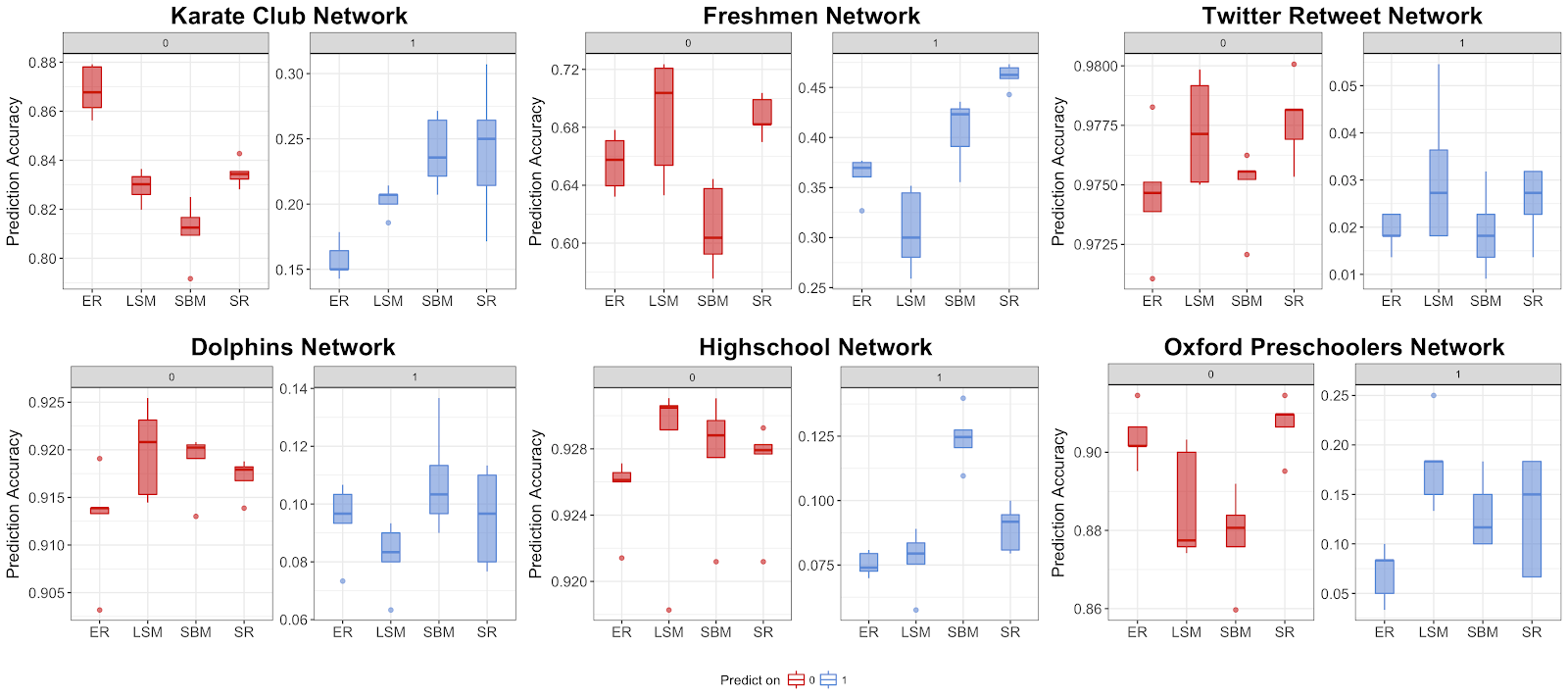}
  \caption{SCV Results on Each Social Network (Red: prediction on non-edges, Blue: prediction on edges)}
\end{figure}

The top leftmost plot in Figure 8 displays the SCV results for the Karate Club Network. We firstly fit the SBM with both $2$ and $3$ blocks and visualized the block assignments for comparison. It turns out that the SBM with 3 blocks looked relatively more reasonable than than the SBM with $2$ blocks, which did not tell a sensible story about the formations of the relationships. Hence, $3$ was selected as the number of blocks for the SBM in the SCV for this network. The prediction accuracies on edges range from $0.15$ to $0.3$, with those for the SBM and the SR model overlapping with each other. The relatively low prediction accuracies yielded by the ER model suggests the existence of a clear pattern or structure in the network. It is also interesting that the ER model has the highest mean prediction accuracy on non-edges but the lowest on edges. Overall, both the SBM ($3$ blocks) and the SR model work similarly well compared to the other two. However, since the SR model has larger variation, we consider the SBM the winner model.

The Freshmen Network is the only network among the six analyzed in this study in which the density is greater than $0.3$. An interesting result is that the overall gap between the prediction accuracies on the non-edges (red) and those on the edges (blue) is considerably small compared to the other networks, which are much more sparse. This feature indicates that SCV tends to yield better prediction accuracies on edges for denser networks. Another important fact is that unlike most of the networks (except for the Highschool Network), this network has a clear winner according to the prediction accuracy on edges: the SR model. The SR model not only has the highest mean accuracy but also the smallest variation, suggesting that the parameter estimates are stable relative to other models. So compared to block structures (SBM) or similarity among nodes (LSM), personal tendencies of forming friendships with other entities have a huge influence on the formation of all the relationships within this social network.

In the SCV result plot for the Twitter Retweet Network, we observe that the mean prediction accuracy on edges is extremely low. Overall, the boxplots for all four models overlap each other, indicating that there is no significant difference between prediction performance across the models. Even so, we observe that the mean accuracy of the ER model is one of the lowest. This result suggests that the Twitter Retweet Network has a clearly defined network structure. The SBM fit on $2$ blocks also seems to perform as bad in terms of mean prediction accuracy, with larger variance and wider range. From this we can infer that there might not be a conspicuous community or block structure within this network. On the other hand, the LSM with $2$ dimensions and the SR model seem to perform relatively better. Although the mean prediction accuracy for these two models are moderately similar, we must take note that in the simulations, the LSM produced the largest variance. Not only is the spread wide within the interquartile range, but also the longer upper tail implies that the distribution of the prediction accuracies are skewed. 

The bottom leftmost plot in Figure 8 shows the result for the Dolphins Network. The prediction accuracies on edges range from $0.08$ to $0.14$, excluding outliers. Since the LSM has lowest mean prediction accuracy on edges and largest variation for accuracies on non-edges, it fits badly on this network. The relatively high prediction accuracies of the ER model may suggest that there is no clear structure for this network. The SBM and the SR model have similar mean accuracies to the ER model's. Since no model outperforms the ER model, we attempt to conclude that there is no clear structure in the Dolphin Network that the non-ER CID models can capture and produce good fits. But for the sake of selecting the final model, considering the relatively small variation for prediction accuracies on edges and the high prediction accuracies on non-edges of the SBM, we finally decide to fit an SBM ($2$ blocks) on the Dolphins Network.

Just like for the Freshmen Network, there is a clear winner model for the Highschool Network, at least according to the SCV accuracies: the SBM with $3$ blocks. As shown in the plot of the original network (Figure 2), there is a clear triangular structure within the graph, and thus it is not surprising that the $3$-block SBM outperforms the other models. Furthermore, the ER model has the lowest mean accuracy for prediction on edges. Such a result suggests that there is some clear structure within the interactions among the highschool students in this network, and as just mentioned, it turned out that the SBM captures that structure the best among the non-ER models. Another interesting fact is that this network is the only one where the SBM outperforms the SR model in prediction on edges. This network is an exception to the general trend of the SR model performing at least as well as the SBM for most cases, which will be discussed more later on.

For the Oxford Preschoolers Network, in general the prediction accuracies range from $0.1$ to $0.2$, except for the ER model's. For this network, the ER's performance is notably low compared to rest of the models. So, we may infer that structural components in the Oxford Preschoolers Network are considerably clear. Simulations of the SBM were conducted using $2$ blocks since we observed that doing so produces a better fit than using $3$ blocks on the original network. Nevertheless, the SBM does not show high performance when compared to the LSM and the SR model. Although the ranges of the three boxplots for the LSM, the SBM and the SR model overlap, we observe distinct difference in the mean prediction accuracy. The SR model performs second best, but the boxplot displays a large degree of variation which could be an indicator that model fit is actually bad and induces very unstable parameter estimates. The LSM with $2$ dimensions outperforms the rest of the models for the Oxford Preschoolers Network. It not only has the highest mean prediction accuracy, but also has the smallest variation within the computed accuracies although there is an outlier in the upper tail.

\section{Model Fit Results}
\noindent
For the non-ER models (i.e. SBM, SR, and LSM), given the parameter estimates, we display the model fit via the following plots:

\begin{itemize}
\item\underline{SBM}: We set a color for each block, and color each node using the color corresponding to the block it is assigned to. It is expected that nodes with the same colors are somehow clustered so that block divisions are clearly visible.
\item \underline{SR}: We draw two plots, where in the first plot the node size is proportional to the fitted sender parameters, and in the second plot the node size is proportional to the fitted receiver parameters. It is expected that each node has node size reasonably proportional to the number of incoming edges (receivers plot) or that of outgoing edges (senders plot).
\item \underline{LSM ($2$-dimensional)}: We plot the latent space positions in two dimensions, and add a line between every pair of points that correspond to a pair of nodes that have an edge. It is expected that points close to each other in the latent space tend to be connected, while points far away are not.
\end{itemize}
Based on the SCV results displayed in the boxplots, we chose one winner model for each network. The final model fit results are shown in the following.

\subsection{SBM (3 Blocks) Fit on Highschool Network}
\noindent
According to Figure 2, the Highschool Network has a reasonably clear cluster structure. The entire network has a shape similar to a right triangle, where nodes located near each of the three points of the triangle naturally form a community. Together with the SCV result of the SBM with $3$ blocks being the clear winner, we decided the $3$-block SBM as the final model for this network.

Unsurprisingly, an SBM with $k = 3$ blocks was a good choice: the fitted block assignments are similar to what we expected when we visually saw the original data, as shown in Figure 9 below. The three communities are formed near the three points of the triangle that we thought of, and more importantly, nodes in the same blocks are connected much more frequently compared to nodes in different blocks. One interesting result is that most of the 70 probabilistic block assignments are of probability $1$, so the block memberships were assigned with great levels of certainty.

\begin{figure}[h!]
\centering
  \includegraphics[width=6cm, height=6cm]{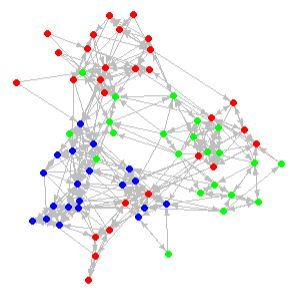}
  \caption{Block Assignments ($k = 3$) on Highschool Network}
\end{figure}

\subsection{SBM (3 Blocks) Fit on Karate Club Network}
\noindent
As mentioned in Section 6, the choice of $3$ as the number of blocks in the SBM for this network was only because the $3$-block fit looked \textit{relatively} better than the $2$-block fit, not because the former captured clear clusters. The visualization denoted by $3$ different colors (Figure 10) does show a cluster pattern at the upper right corner (denoted by blue nodes); however, we fail to detect cluster patterns for pink and green nodes.

We attempt to conclude that, based on the model fitting plots and block assignments obtained so far, the SBM with $3$ blocks is not suitable for identifying clusters for the Karate Club Network. Instead, it identifies club members who are in the core of this network from those who are in the periphery. For example, within-group edge density for group 1 (pink nodes), group 2 (blue nodes) and group 3 (green nodes) are 1, 0.25 and 0.077, respectively. It is also interesting to look at the edge probabilities. In this $3$-Block SBM, the within-group edge probability for group 1 (1.00) is the highest, and that for group 2 and group 3 respectively decreases to 0.293 and 0.099. The lowest between-group edge probability exists between group 2 and group 3, $0.013$; while the highest between-group edge probability is for group 1 and group 3, with a probability of $0.318$. These quantitative measures well present the plot structure but with higher accuracy than visual estimation.

\begin{figure}[h!]
\centering
  \includegraphics[width=6cm, height=6cm]{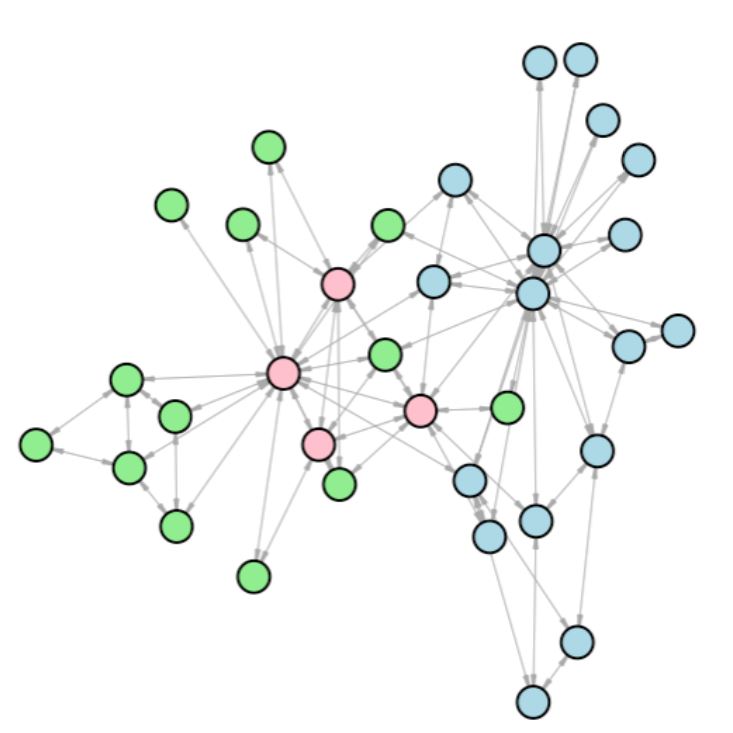}
  \caption{Block Assignments ($k = 3$) on Karate Club Network}
\end{figure}

\subsection{LSM (2 Dimensions) Fit on Oxford Preschoolers Network}
\noindent
After fitting an LSM with $2$ dimensions on the Oxford Preschoolers Network, we plotted the obtained models parameters: positions on dimension 1 and dimension 2. To enhance interpretability, arrows indicating directed edges in the original network were added to the plot (Figure 11). 

One eye-catching trait in Figure 11 is that nodes placed close on the latent space tend to be actually connected in the original network. On the bottom right corner of the plot, we observe a cluster of nodes placed close to each other. Toward this corner of the plot, there is an increase in the number of observed edges. However, as nodes get placed further away from this corner, the number of edges decreases. Two nodes that have no connections in the network are indeed located furthest away from the rest of the preschoolers. Therefore, the LSM captures this network structure well and thus produces good model fit results. 

\begin{figure}[h!]
\centering
  \includegraphics[width=8cm, height=8cm]{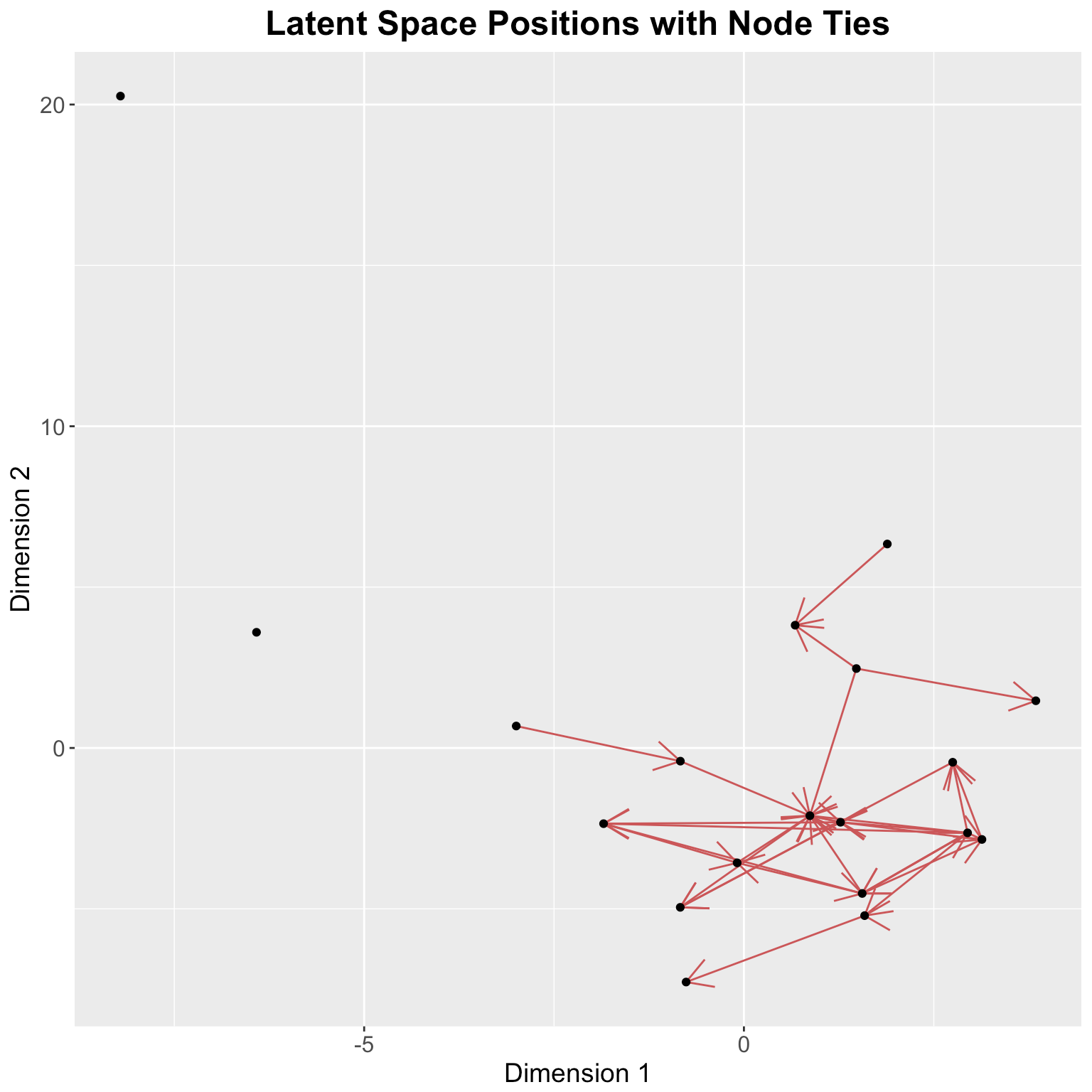}
  \caption{Latent Space Positions for Oxford Preschoolers Network}
\end{figure}

\subsection{SR Fit on Freshmen Network}
\noindent
Figure 12 shows the two plots corresponding to the fit of the SR model on the Freshmen Network. A structural feature of this network is that most of the nodes are highly interconnected among one another at the central part of the plot, and the remaining few nodes such as Node 9 and Node 26 are in the outer boundary. Those few nodes in the outer boundary do not have much outgoing edges compared to the central nodes, meaning that they are less outgoing than the majority of the entities.

According to Figure 12, the sender parameter estimates of the central nodes are considerably large compared to the few outer nodes. On the other hand, the relative gap of the receiver parameter estimates is very small. These plots suggest that the central nodes are much more active in sending (i.e. request social relationships), while both the central and outer nodes are similarly active in receiving such requests. This implication can indeed be verified by the degrees of each node, and thus the SR model successfully conveys a story behind the formation of the interactions in this social network. Such a result is consistent to our common sense in the real world as well. This network deals with college freshmen in the same department, so it has not been long since these students met one another in school. Common sense indicates that at such an early time point, individual tendency of initially forming social relationships (which an SR model seeks) is more influential than clique structures among the students (which an SBM seeks) and common interests or similarities (which an LSM seeks).

\begin{figure}[h!]
\centering
  \includegraphics[width=12cm, height=6cm]{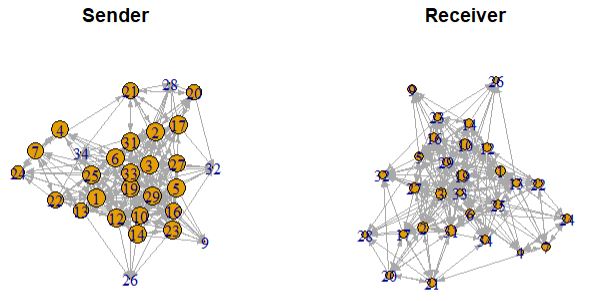}
  \caption{Sender and Receiver Parameter Estimate Plots for Freshmen Network}
\end{figure}

\subsection{SBM (2 Blocks) Fit on Dolphins Network}
\noindent
As mentioned in section 6, we fit an SBM with $2$ blocks on the Dolphins Network. According to the initial plot, there is a two-cluster pattern in the Dolphins Network. When we fit a $2$-block SBM on the Dolphins Network, not surprisingly, the visualization (Figure 13) where the $2$ colors represent the block assignments for each node tells a sensible story about the distribution of blocks; namely, there does exist an obvious display of two clusters just as what we have assumed. In addition, the block assignment probabilities are mostly near $1$ so the model has a high confidence in choosing one group over the other one for each node and is highly confident in its model fit. We can therefore conclude that the SBM with $2$ blocks is a good model to fit on the Dolphins Network. 

\begin{figure}[h!]
\centering
  \includegraphics[width=6cm, height=6cm]{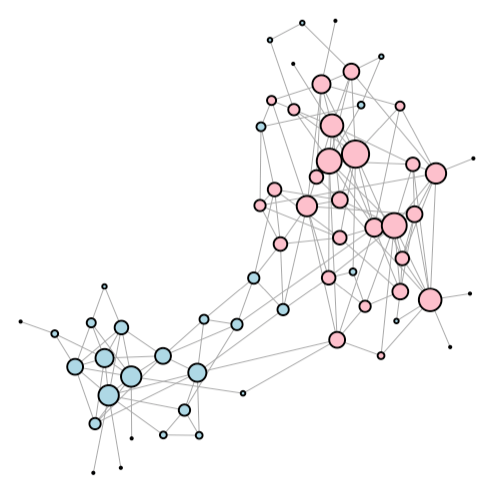}
  \caption{Block Assignments ($k = 2$) on for Dolphins Network}
\end{figure}

\subsection{LSM (2 Dimensions) Fit on Twitter Retweet Network}
\noindent
Figure 14 shows the estimated model parameters, which are the positions of the nodes in the $2$-dimensional Euclidean space, together with lines between nodes that have an edge. Since the Twitter Retweet Network is an undirected graph, each edge in the original network was represented as lines, not arrows, on the latent space. 

Similarly to Figure 11, Figure 14 shows that nodes positioned close on the latent space were indeed connected in the original graph. This enforces the transitive relationship between the nodes. Each edge represents an instance of a retweet, for which we expect to occur between posts of similar content. Therefore, it is reasonable that users who posted similar content are placed closer on the latent space and are connected to each other since it becomes more likely form them to retweet each other's posts. Another feature we observe is that there seems to be clusters of nodes that connect to a central node. We conjecture that this represents an influential popular user at the center and several followers who share the same interest at its periphery. Moreover, node connectivity increases at the center of Figure 14, and decreases toward the outskirts of the plot.

\begin{figure}[h!]
\centering
  \includegraphics[width=8cm, height=8cm]{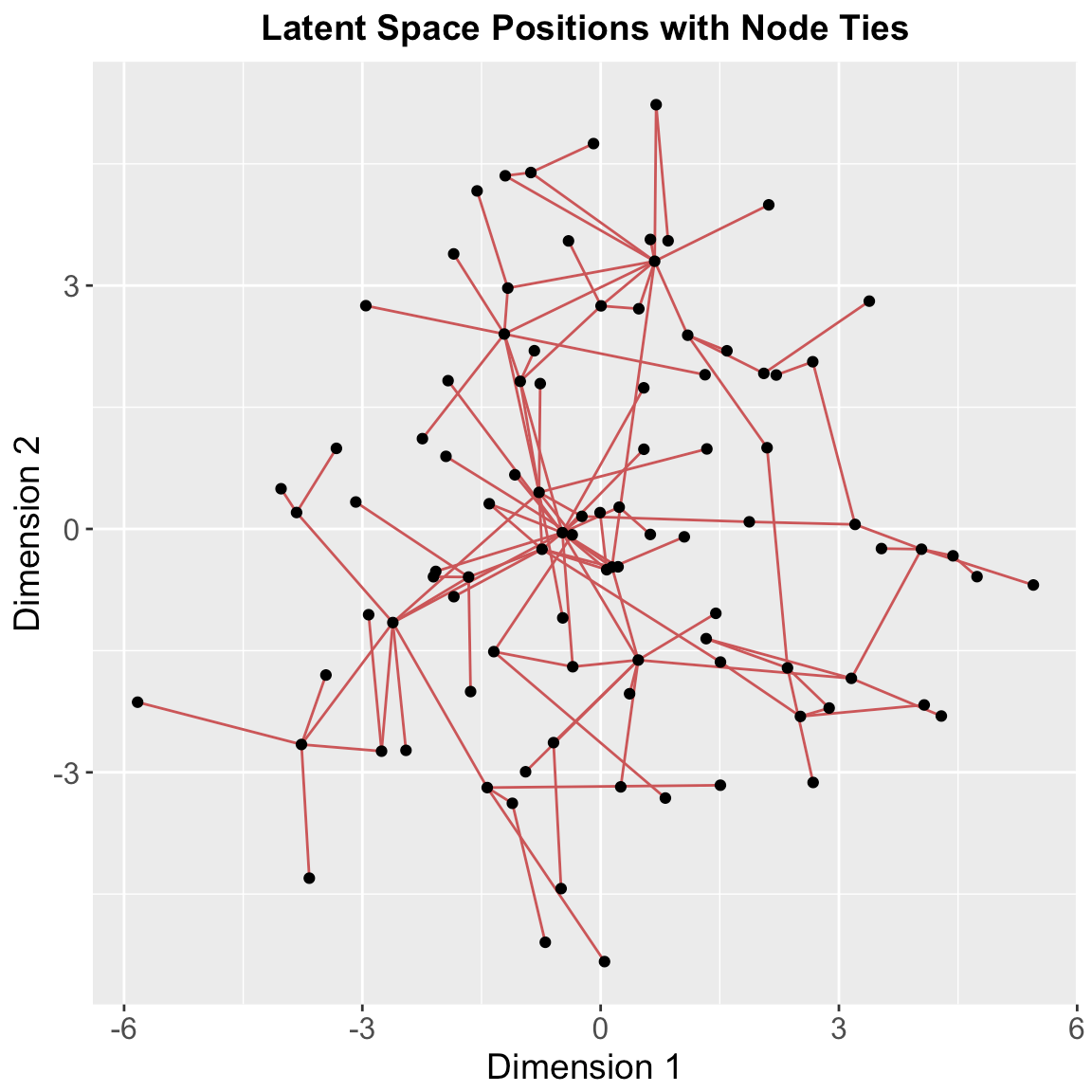}
  \caption{Latent Space Positions for Twitter Retweet Network}
\end{figure}

\section{Conclusions and Discussions}
\noindent
Based on our SCV procedures and final model fit results, we conclude the following major findings. 

Considering how social relationships existent in a network were generated, the more there were clear patterns or factors that affected each relationship, the less it is likely the ER model performs well. That is, a model would produce a good fit on structured networks only if it is capable of capturing that pattern that generated the edges between nodes. For example, in the Highschool Network, the SBM was able to capture the three-block structure that led to the formations of the edges in the graph reasonably well. However, the ER model assumes that there is no such pattern in this network. Instead, it argues, or hopes, that every edge is simply a result of a random coin toss with a constant probability of heads, and thus it cannot capture the significant structures, if any, in a network. This result is verified by the SCV accuracies (Figure 8); there is no network such that the ER model and another non-ER model are the two winners. So if there is a non-ER model, which seeks to capture some type of structure (depending on the model definition), that performs reasonably well, then it is very unlikely that the ER model will do similarly good.

Another interesting pattern is that for most networks, the prediction accuracies on edges of the SR model are at least as good as those of the SBM. The only exception is the Highschool Network in which the SBM outperformed all other models. Recall that the SR model captures the individual tendency of sending and receiving requests for social relationships, and the SBM seeks to find an optimal group assignment for each of the entities. Note that the formations of groups among people are influenced by the personal tendencies in the first place. For example, if most of the people are outgoing (sender) and/or popular (receiver), it is more likely that groups will be actively formed. Therefore, we observe this pattern that if capturing personal tendencies can be done well, then so can capturing group formations among the entities.

Moreover, the more clear the definition of 'similarity' of entities is, the higher the LSM's prediction accuracies on edges tend to be. The LSM tries to map each node to a latent space position in such a way that the distance between two nodes is inversely proportional to how likely they are connected. Therefore, the model needs a clear notion of 'similarity' between nodes that are connected by an edge to produce a good fit. In this study, two networks had good LSM fits: the Oxford Preschoolers Network and the Twitter Retweet Network. In the former, similarity can be represented as the common interest towards the same toy. That is: the fact that two nodes have a connecting edge means that the two kids competed for a single toy, which implies that both of them had a desire for the same toy. In the Twitter Retweet Network, the notion of similarity is even more explicit. The existence of an edge between two nodes implies that the corresponding to Twitter users retweeted each other's posts, meaning that they have some common interest or thought about an issue.

We also observed two major limitations in our self-developed GOF measure. The first is that the SCV tends to yield higher prediction accuracies on edges when the network is more dense. This characteristic is not surprising since higher density implies that the model has more information about how the existent edges were formulated. However, most real-world social networks are (extremely) sparse. Realistically speaking, considering the numerous entities around a person, it is difficult for him or her to initialize and maintain social interactions with all (or the majority) of them. Therefore, in order for the SCV metric to be applied well in real-world social network datasets, it should be improved so that it works well even on sparse networks.

The second limitation is the variability in the SCV results. The boxplots in Figure 8 show that some model fits have very large difference in the mean prediction accuracy across different trials. One conjecture about this trend is that the particular model chosen for a certain network is not the optimal one and thus the fits themselves turn out to be unstable. In other words, the better the model fits on the network, the more stable the estimation on the parameters and the SCV results will be.

\section{Future Work}
\noindent
Based on the limitations of our study, we propose the following directions for future work. The first is to try more models and also more networks. For example, we chose the SBM with $3$ blocks as the final model to fit on the Karate Club Network. However, we can hardly conclude that the SBM is the best model. Rather, the reason why we chose this model is that it works better than other three models considering the higher variation and lower prediction accuracies of the results of other models. Trying more types of models, or even trying different hyperparameter values for the models already explored in this study, can help find a better one. The \textsf{CIDnetworks} R package$^{13}$ currently supports the following other models: the covariate model$^{14}$, the hierarchical block model$^{15}$, the latent vector model$^{16}$ and the mixed membership stochastic block model$^{17}$. Another possible model is the degree-corrected stochastic block model$^{18}$, which is a combination of the SBM and the SR model. Similarly, performing these comparisons on more real-world social networks would give us much more insights. In this study, we analyzed six networks and did find some interesting patterns. But in terms of generalizing such patterns, results on more social networks would be helpful. Appendix A contains the network data descriptions of the four additional real-world social networks which we have explored but haven't applied the SCV and model-fitting yet. 

Another possible approach is to incorporate generative process in our GOF measure. Sample networks can be generated via the estimated parameter values from the best model based on the SCV accuracies. More specifically, for each pair of nodes in the original network, we compute the estimated edge probability using the parameter estimates and the formula defined by the model, and then run a Bernoulli trial to decide whether or not to put an edge. But if such newly-generated networks are considerably different from the original network, then either the model or the SCV itself as metric is problematic. Utilizing these generative capabilities could provide more insights. Note that we must specify a criterion of the difference between two networks in order to use this approach.

Finally, we suggest identifying appropriate prior distributions used in Bayesian inference for each network involved in estimation process. The \textsf{CIDnetworks} R package performs Bayesian estimation on model parameters and requires a prior distribution to be pre-specified. Using Gibbs sampling, the \textit{CID.Gibbs} function in the package performs Markov Chain Monte Carlo that generates a sample from the distribution that is expected to be close to that of the parameters and takes the mean as the parameter estimate. In our study, we mainly used the default prior distributions given in the package. We believe that finding more network-specific prior distributions will improve the estimation process and yield better model fit quality. In addition, we firmly believe in the need to lift the computational constraint on the time-consuming simulation process. Optimizing the current source code in the \textsf{CIDnetworks} R package can make the model fitting process more efficient and quick so that we can examine multiple trials within a smaller time window.



\newpage

\section*{Acknowledgements}
\noindent
We sincerely thank Dr. Brian Junker and Dr. Nynke Niezink in the Department of Statistics \& Data Science at Carnegie Mellon University for their guidance on the direction of this research project, their effort to improve our working progress, their time for holding weekly meetings with us and finally, their unconditional support and encouragement. We also express our gratitude to Dr. Tracy Sweet in the Department of Human Development and Quantitative Methodology at University of Maryland for providing the learning materials about the software used in this project.

This work was supported in part by the US Department of Education, Institute of Education Sciences, under grant \#R305D150045. Any opinions, findings, and conclusions or recommendations expressed in this material are those of the author(s) and do not necessarily reflect the views of the granting agencies.


\newpage
\section*{Appendix A}
\noindent
This appendix contains the network data descriptions of the four additional real-world social networks that we have explored but have not applied SCV and model-fitting yet. Table 2 shows the summary statistics of these four social networks.

\begin{table}[htbp]
\centering
\begin{tabular}{ccccc}
\hline
Network & Number of Nodes & Number of Edges & Density & Reciprocity \\
\hline
\hline
Enron Network & 151 & 266 & 0.012 & 0.075\\
Students Cooperation Network & 185 & 360 & 0.021 & 1 \\
Residence Hall Network & 209 & 900 & 0.0207 & 0.529 \\
Divorce Network & 59 & 225 & 0.132 & 1 \\
\hline
\end{tabular}
\caption{Summary Statistics of Each Social Network in Appendix A}
\end{table}

\subsection*{1) Enron Network}
\noindent
The Enron Network data$^{19}$ describe two types of social relationships among $151$ people who work in Enron company: manager-subordinate relationship and colleague relationship. There are total $266$ edges (i.e. relationships), where exactly half ($133$ edges) have the manager-subordinate relationship type, and the other $133$ edges have the colleague relationship type. The edges are directed specifically because of the nature of manager-subordinate relationships. Note that each colleague edge is also expressed as a directed edge, but we should interpret it as an undirected edge since a colleague relationship is naturally symmetric. For some reason, as depicted in the reciprocity, around $7.5$\% of the edges are bi-directional. Figure 15 shows a plot of the network.

\begin{figure}[h!]
\centering
  \includegraphics[width=7.5cm, height=6.5cm]{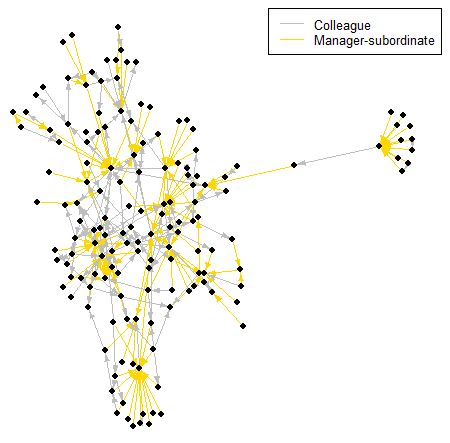}
  \caption{Enron Network}
\end{figure}

\subsection*{2) Students Cooperation Network}
\noindent
The Students Cooperation Network data$^{20}$ were collected by the BGU (Ben-Gurion University) Social Networks Security Research Group. This social network contains information about $185$ participating students from two different departments. Each undirected edge represents a cooperation of the corresponding pair of students while they were working on there homework assignments. The network contains total $360$ edges, and each edge has one of the three types: ``Partners'', ``Computer'', and ``Time''. A ``Partners'' edge represents the explicit connection between the students who submitted theoretical or coding assignments together as partners. A ``Computer'' edge is the implicit connection between students who used the same computer for solving an online assignment. Lastly, a ``Time'' edge is defined as the second implicit connection between students who probably solved the homework assignments together but submitted from different computers. Figure 16 shows a plot of the network.

\begin{figure}[h!]
\centering
  \includegraphics[width=9cm, height=9cm]{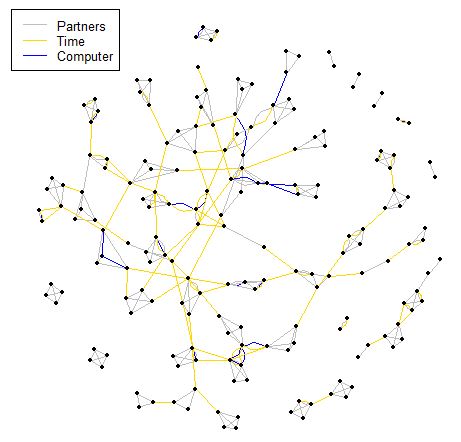}
  \caption{Students Cooperation Network}
\end{figure}

\subsection*{3) Residence Hall Network}
\noindent
The Residence Hall Network data$^{21}$ describes friendship relationships among $217$ students at a residence hall located on the Australian National University campus. This social network contains $2672$ edges each representing a friendship. Every edge is directed and has a weight value $\in \{1, \cdots , 5 \}$, which is proportional to the strength of the friendship. Due to such a large number of edges, if we were to visualize the original network, it will be extremely difficult to see the structures in the network. Therefore, we only plot the edges with weights of $4$ or $5$ since they represent more intimate friendships compared to the other edges. The term ``Residence Hall Network'' (including those in Table 2 and Figure 17) refers to this subnetwork, and not the original network. Figure 17 shows a plot of the subnetwork.

\begin{figure}[h!]
\centering
  \includegraphics[width=7.5cm, height=8.5cm]{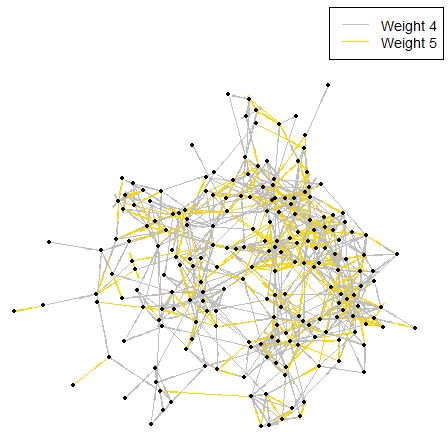}
  \caption{Residence Hall Network}
\end{figure}

\subsection*{4) Divorce Network}
\noindent
The Divorce Network data$^{22}$ contains information about $9$ types of divorce laws in the $50$ US states. This network is a legal basis for divorce in the states and is the only bipartite graph among the networks introduced in this paper. That is: the network contains a set of $50$ nodes corresponding to the states and another of $9$ nodes each representing a type of divorce law. Moreover, the $225$ unweighted edges, each corresponding to a pair of a state and a category, indicate that the state constitutes a divorce law in that category. The $9$ types of divorce laws included in this network are: ``incompat'' (incompatible of temperament), ``cruelty'' (cruelty), ``desertn'' (disertion), ``nonsupp'' (non-supplement), ``alcohol'' (alcohol), ``felony'' (felony), ``impotenc'' (impotence), ``insanity'' (insanity) and ``separate'' (separation). Figure 18 shows a plot of this bipartite network. Since it's hard to see all the edges precisely, we have also included Tables 3 and 4 that show which of the laws each state has.

\begin{figure}[h!]
\centering
  \includegraphics[width=17cm, height=20cm]{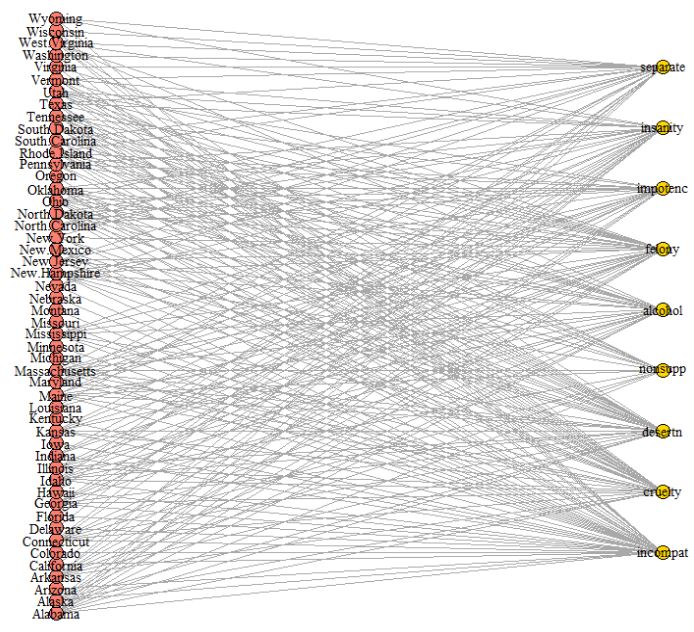}
  \caption{Divorce Network}
\end{figure}

\begin{table}[htbp]
\centering
\begin{tabular}{cccccccccc}
\hline
State&incompat&cruelty&desertn&nonsupp&alcohol&felony&impotenc&insanity&separate \\
\hline
\hline
Alabama &Yes &Yes &Yes &Yes &Yes &Yes &Yes &Yes &Yes \\
Alaska &Yes &Yes &Yes &No &Yes &Yes &Yes &Yes & No \\
Arizona &Yes &No &No &No &No &No &No &No &No \\
Arkansas &No& Yes &Yes &Yes &Yes &Yes &Yes &Yes &Yes \\
California &Yes &No& No& No &No &No& No &Yes& No \\
Colorado &Yes& No& No &No &No &No& No& No& No \\
Connecticut& Yes& Yes &Yes &Yes& Yes& Yes &No &Yes &Yes \\
Delaware &Yes &No &No &No& No& No &No& No& Yes \\
Florida& Yes& No& No& No &No &No& No& Yes &No \\
Georgia& Yes &Yes &Yes &No& Yes& Yes& Yes &Yes& No \\
Hawaii &Yes& No& No& No& No& No& No &No &Yes \\
Idaho& Yes &Yes &Yes &Yes& Yes& Yes& No& Yes& Yes \\
Illinois &No &Yes& Yes& No& Yes &Yes& Yes& No& No \\
Indiana &Yes& No& No& No &No &Yes &Yes &Yes &No \\
Iowa &Yes &No &No& No &No& No &No& No& No \\
Kansas &Yes& Yes &Yes& No &Yes &Yes &Yes &Yes& No \\
Kentucky &Yes &No &No& No& No &No &No &No &No \\
Louisiana &No &No& No& No &No &Yes &No &No &Yes \\
Maine &Yes &Yes &Yes &Yes &Yes &No &Yes &Yes& No \\
Maryland &No &Yes &Yes& No& No &Yes &Yes &Yes &Yes \\
Massachusetts &Yes &Yes &Yes &Yes &Yes& Yes &Yes& No& Yes \\
Michigan &Yes &No& No& No& No &No &No &No &No \\
Minnesota &Yes& No& No &No &No &No& No &No &No \\
Mississippi &Yes& Yes &Yes &No &Yes &Yes &Yes &Yes& No \\
Missouri &Yes &No &No &No &No &No& No& No& No \\
\hline
\end{tabular}
\caption{Types of Laws that Each State Has in Divorce Network (1/2)}
\end{table}

\begin{table}[htbp]
\centering
\begin{tabular}{cccccccccc}
\hline
State&incompat&cruelty&desertn&nonsupp&alcohol&felony&impotenc&insanity&separate \\
\hline
\hline
Montana &Yes &No &No &No &No &No &No &No &No \\
Nebraska &Yes& No &No &No &No &No& No& No& No \\
Nevada &Yes &No &No &No &No &No &No &Yes &Yes \\
New.Hampshire &Yes& Yes &Yes &Yes &Yes &Yes &Yes &No &No \\
New.Jersey& No& Yes& Yes& No &Yes &Yes &No& Yes &Yes \\
New.Mexico &Yes& Yes& Yes& No& No& No &No &No &No \\
New.York &No& Yes &Yes& No& No& Yes& No &No& Yes \\
North.Carolina& No& No& No &No &No& No& Yes &Yes &Yes \\
North.Dakota &Yes &Yes &Yes &Yes &Yes& Yes& Yes& Yes& No \\
Ohio& Yes& Yes &Yes &No &Yes& Yes &Yes &No& Yes \\
Oklahoma& Yes &Yes &Yes &Yes &Yes& Yes& Yes &Yes& No \\
Oregon &Yes& No &No &No& No& No& No &No &No \\
Pennsylvania &No& Yes &Yes& No &No& Yes& Yes& Yes& No \\
Rhode.Island& Yes& Yes& Yes &Yes &Yes& Yes& Yes &No &Yes \\
South.Carolina &No& Yes& Yes& No &Yes& No& No &No& Yes \\
South.Dakota& No &Yes& Yes &Yes &Yes& Yes &No& No& No \\
Tennessee &Yes& Yes &Yes &Yes &Yes& Yes &Yes& No& No \\
Texas& Yes& Yes &Yes &No &No &Yes& No &Yes &Yes \\
Utah& No& Yes& Yes& Yes& Yes& Yes& Yes &Yes& No \\
Vermont& No &Yes& Yes &Yes &No &Yes& No& Yes& Yes \\
Virginia &No& Yes& No& No &No& Yes& No& No& Yes \\
Washington& Yes& No& No& No& No &No& No& No& Yes \\
West.Virginia& Yes& Yes& Yes& No& Yes& Yes &No &Yes& Yes \\
Wisconsin &Yes &No &No &No &No &No &No& No& Yes \\
Wyoming &Yes &No & No& No &No& No& No& Yes& Yes \\
\hline
\end{tabular}
\caption{Types of Laws that Each State Has in Divorce Network (2/2)}
\end{table}

\end{document}